\documentclass[graybox]{svmult}

\usepackage{mathptmx}       
\usepackage{helvet}         
\usepackage{courier}        
\usepackage{type1cm}        
\usepackage{makeidx}         
\usepackage{graphicx}        
\usepackage{multicol}        
\usepackage[bottom]{footmisc}

\makeindex             

\begin{document}

\title*{Predatory trading and risk minimisation: how to (b)eat the competition}
\author{Anita Mehta}
\institute{S. N. Bose National Centre for Basic Sciences,  Block-JD Sector III, Salt Lake, Calcutta-700098, India, \email{anita@bose.res.in}}
%
%
\maketitle

\abstract{ We present a model of predatory traders interacting with each other in the presence of a central reserve (which dissipates their wealth through say, taxation), as well as inflation. This model is examined on a network for the purposes of correlating complexity of interactions with systemic risk. We suggest the use of selective networking to enhance the survival rates of arbitrarily chosen traders. Our conclusions show that networking with 'doomed' traders is the most risk-free scenario, and that if a trader is to network with peers, it is far better to do so with those who have less intrinsic wealth than himself to ensure individual, and perhaps systemic stability.}

\section{Introduction}
\label{sec:1}

The topic of predatory trading and its links with systemic risk is of great contemporary interest: at the time of writing this paper, these links have been mentioned repeatedly in the World Economic Forum at Davos, in addition to having formed the backbone of the 'Occupy' movements worldwide. Immense public anger has been expressed against corporate greed (with predatory trading forming a major way that this is manifested), and many intellectuals worldwide attribute this to the collapse of the world economic system. In this paper, we examine these ideas in a more technical way to see if rigorous mathematical links can be established between these two concepts.

In order to put our mathematical models in the context of current interdisciplinary literature, we quote the conclusions of two key papers. First we define predatory trading along the lines of a recent paper  \cite{bp}, as that which induces and/or exploits other investors' need to 'reduce' their positions. If one trader needs to sell, others also sell and subsequently buy back the asset, which leads to price overshooting and a reduced liquidation value for the distressed trader. In this way,  a trader profits from triggering another trader's crisis; according to the authors  of  \cite{bp}, the crisis can spill over 'across traders and across markets'. To model this scenario, we invoke a model of predatory traders in the presence of a central reserve \cite{am_wealth}, which is principally a source of wealth dissipation in the form of taxation. We assume that this dissipation acts uniformly across the traders' wealth, irrespective of their actual magnitudes. Among our findings \cite{am_wealth} is the fact that when all traders are interconnected and interacting, the entire system collapses, with one or zero survivors. This finds resonance with the ideas of another key paper \cite{haldane}, where analogies with model ecosystems have led the authors to conclude that propagating complexity (via the increase of the {\it number} and {\it strength} of interactions between different units) can jeopardise systemic stability. 

The original model \cite{luck_am} on which \cite{am_wealth} is based, was introduced as a model of complexity. It embodies predator-prey interactions, but goes beyond  the best-known predator-prey model due to Lotka and Volterra by embedding interacting traders in an active medium; this is a case where the Lotka-Volterra model cannot be simply applied.  As mentioned above, a central reserve bank represents such an active medium in the case of interacting traders, whose global role is to reduce the value of held wealth as a function of time \cite{am_wealth}. This forms a more realistic social backdrop to the phenomenon of predatory trading, and it is this model that we study in this paper. In order to relate it to the phenomenon of systemic risk, we embed the model on complex networks \cite{watts,albert}; these represent a compromise between the unrealistic extremes of mean field, where all traders interact with all others (too global) and lattice models, where interaction is confined to  local neighbourhoods (too local). Many real world networks, in spite of their inherent differences,  have been found to have the topology of complex networks \cite{rev_comnets, soc_nets}; and the embedding of our model on such networks \cite{nirmal} allows us to probe the relevance of predatory trading to systemic stability.

The plan of the paper is as follows. First, in Sec. \ref{sec_model}, we introduce the model  of interacting traders of varying wealth in the presence of a central reserve, and show that typically only the wealthiest survive. Next, in Sec. \ref{sec_compnet}, we probe the effect of networks: starting with an existing lattice of interacting traders with nearest-neighbour interactions,  we add non-local links between them with probability $p$ \cite{watts}.  Survivor ratios are then measured as a function of this 'wiring probability' $p$, as $p$ is increased to reflect the topologies of small world and fully random networks  ($p=1$). In Sec. \ref{networksmall}, we ask the following question: can the destiny of a selected trader be changed by suitable networking?  We probe this systematically by networking a given trader non-locally with others of less, equal and greater wealth and find indeed that a trader who would die in his original  neighbourhood, \textit{is able to change his fate}, becoming a survivor via such selective non-local networking. In Sec. \ref{univ}, we provide a useful statistical measure of survival, the pairwise  probability for a trader to survive against wealthier neighbours, i.e. to win against the odds.  Finally, we discuss the implications of these results to systemic stability in Sec. \ref{discuss}.

\section{Model\label{sec_model}}
The  present model was first used in the context of cosmology to describe the  accretion of  black holes in the presence of a radiation field \cite{archan}. Its applications, however, are considerably more general; used in the context of economics \cite{am_wealth}, it manifests an interesting rich-get-richer behaviour. Here, we review some of its principal properties \cite{luck_am}.

Consider an array of traders with time-dependent wealth $m_i (t)$ located at the sites of a regular lattice. The time evolution of wealth of the traders is given by the coupled deterministic first order equations,
\begin{eqnarray}
\frac{dm_i}{dt} &=& \left(\frac{\alpha}{t} - \frac{1}{t^{1/2}} \sum_{j \neq i} g_{ij} \frac{dm_j}{dt} \right) m_i  - \frac{1}{m_i}.
\label{blackhole}
\end{eqnarray}

Here, the parameter $\alpha$ is called the wealth accretion parameter (modelling investments, savings etc) and $g_{ij}$ defines the strength of the interaction between the traders $m_i$ and $m_j$. The first parenthesis in the R. H. S of  Eqn. \ref{blackhole} represents the wealth gain of the $i$th trader, which has two components: his wealth gain due to investments/savings (proportional to $\alpha$) modulated by dissipation (at the rate of $1/t$) due to e.g. taxation, and his wealth gain due to predatory trading (the second term of Eqn. \ref{blackhole}), also modulated by dissipation (at the rate of $1/{t^{1/2}}$) in the same way. Notice in the second term, that the loss of the other traders corresponds to the gain of the $i$th trader, so that each trader 'feeds off' the others, thus justifying the name 'predatory trading'.  The last term, $-1/m_i$, represents the loss of the $i$th trader's wealth through inflation to the surroundings; we will see that this term ensures that those without a threshold level of wealth 'perish', as in the case of those individuals in society who live below the poverty line. Here and in the following we will use words such as 'dying' or 'perishing' to connote the bankruptcy/impoverishment, of a trader and conversely, 'life' will be associated with {\it financial} survival, i.e. solvency.

A logarithmic time is introduced in the  study for convenience. We define a scaled time $s = \ln(t/t_0)$, where $t_0$ is some initial time. Similarly, for convenience, we rescale wealth to be  $X_i = m_i/t^{1/2}$. Using the new variables, Eqn. \ref{blackhole} can be rewritten as,
\begin{eqnarray}
\frac{dX_i}{ds} \equiv {X'}_i = \left(\frac{2 \alpha -1}{2} - \sum_{j \neq i} g_{ij} \left(\frac{X_j}{2} + {X'}_j  \right) \right) X_i  - \frac{1}{X_i},
\label{model}
\end{eqnarray}
where the primes denote differentiation performed with respect to $s$.

\begin{figure}[!t]
\centering
\resizebox{90mm}{70mm}{\includegraphics{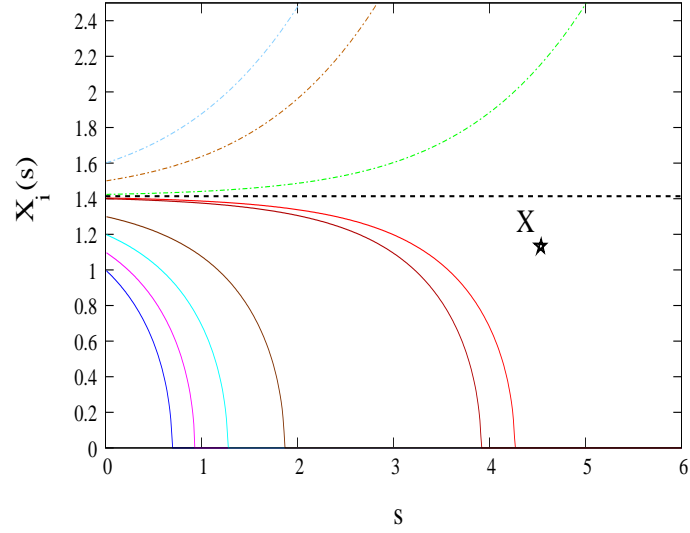}}
\caption{(color online) The plots show the evolution of individual non-interacting traders with a range of initial wealth obeying Eqn. \ref{model}. Traders with initial wealth $X_0$  greater than $X_{\star}$ live forever, or else they die in time. Here $\alpha = 1.0$, therefore $X_{\star} = \sqrt{\frac{2}{2 \alpha-1}} = \sqrt{2}$. \label{freewealth}}
\end{figure}

Continuing our recapitulation of the results of the model \cite{luck_am}, we consider a scenario where there is a single isolated trader, whose initial ('inherited') wealth is $X_0$. Under the dynamics defined by Eqn. \ref{model}, the trader will survive financially only if $X_0>X_{\star} (= \sqrt{\frac{2}{2 \alpha-1}})$ (this imposes the condition $\alpha > 1/2$ \cite{luck_am}), else he will eventually go bankrupt (see Fig. \ref{freewealth}). Next, consider a system of two traders with equal  initial wealth; here, there exists a critical coupling $g_c$ such that for $g<g_c$ the two traders both survive, provided that their individual inherited wealth is greater than $X_{\star}$. For two unequally wealthy traders ($X_1 < X_2$, say), the poorer trader goes bankrupt first at time $s_1$;  the richer one  either survives (if his wealth at time $s_1$,  $X_2(s_1)$, exceeds the threshold $X_{\star}$) or goes bankrupt (if $X_2(s_1) < X_{\star}$). The main inferences are twofold: the wealthier predatory trader 'consumes' the poorer one's wealth in due course, and
then survives or not depending on whether his own wealth at that point is enough to tide him through its eventual dissipation through taxes and inflation. We thus see that this relatively simple model captures not just the mechanism of predatory trading, but also includes the flavour of more sophisticated concepts like inherited wealth, taxation and inflation.

Consider the limit of an infinitely large number of traders all connected to each other; this represents a limiting mean field regime, with fully collective behaviour involving long-range interactions.  For $g>g_c$, all but the wealthiest will eventually go bankrupt. In the weak coupling regime ($g<g_c$) on the other hand, the dynamics consist of two successive stages \cite{luck_am}. In Stage I, the traders behave as if they were isolated from each other (but still in the presence of the reserve);  they get richer (or go bankrupt ) quickly if their individual wealth is greater (or less) than the threshold  $X_{\star}$. In Stage II, slow,  collective and predatory dynamics leads to a scenario where again, only the single wealthiest trader survives. This weakly interacting mean field regime shows the presence of two well-separated time scales, a characteristic feature of  glassy systems \cite{book}. The separation into two stages embodies an interesting physical/sociological scenario: the first stage is fast, and each trader survives or 'dies' only on the basis of his inherited wealth, so that everyone without this threshold wealth is already eliminated before the second stage sets in. Competition and predatoriness enter only in the second stage, when the wealthier feast off their poorer competitors progressively, until there is only overlord left. This is a perfect embodiment of systemic risk, as the entire system collapses, with only one survivor remaining \cite{am_wealth}.

\begin{figure}[!t]
\begin{center}
\resizebox{90mm}{70mm}{\includegraphics{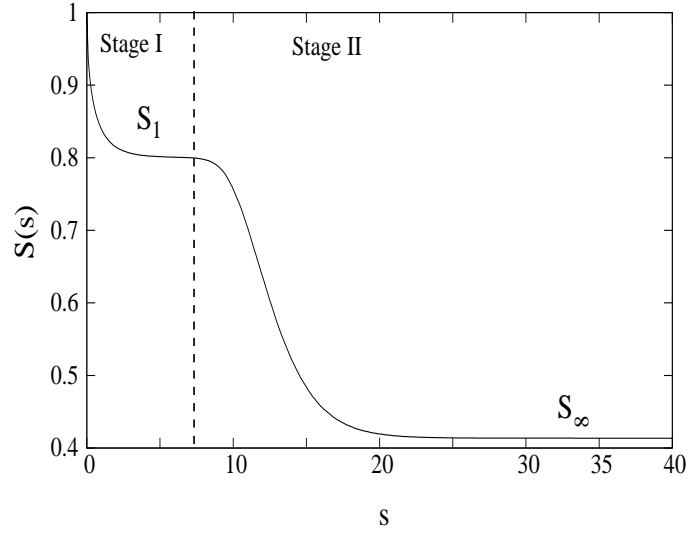}}
\end{center}
\caption{ The survival ratio $S(s)$ plotted as a function of reduced time $s$, for  traders
 distributed in a regular one-dimensional lattice of size $100,000$. In Stage I, traders grow
independently, while in Stage II the growth is collective. Here, $S_1 = 0.8$ is the survival
ratio at the end of Stage I and   $S_{\infty}$ is the asymptotic survival ratio. \label{surratio}}
\end{figure}

Similar glassy  dynamics  also arise  when the model is solved on a periodic lattice with {\it only
nearest-neighbour} interactions. The dynamical equations in  Eqn. \ref{model} take the form :

\begin{eqnarray}
X'_{{\bf n}} &=& \left( \frac{2 \alpha - 1 }{2} + g \sum_{{\bf m}} \left( \frac{1}{X_{\bf m}} - \alpha X_{\bf m} \right) \right) X_{\bf n} - \frac{1}{X_{\bf n}},
\label{finite_model}
\end{eqnarray}
by keeping the terms upto first order in $g$ \cite{luck_am}. Here, ${\bf m}$ runs over the $z$ nearest neighbours of the site ${\bf n}$, where for a 
one-dimensional ring topology, $z=d$, while $z=2d$ for a two-dimensional lattice -- these are the two cases we consider here. We summarize earlier results \cite{luck_am} on the dynamics:  In Stage I, traders evolve independently and (as before) only those whose initial wealth $X_i (s=0)$ exceeds the threshold $X_{\star}$, survive. In Stage II, the dynamics are slow and collective, with competition and predatoriness setting in: however, an important difference with the earlier fully connected case is that there can be {\it several} survivors, provided that these are isolated from each other by defunct or bankrupt traders (i.e.  no competitors remain {\it within their effective domain}). Their number asymptotes to  a constant $S_{\infty}$ (Fig. \ref{surratio}), and these 'isolated overlords' survive forever. The moral of the story is therefore that in the presence of predatory dynamics, interaction-limiting 'firewalls' can help avoid systemic collapse: conversely, {\it full globalisation with predatory dynamics makes systemic collapse inevitable}. Apart from providing quantitative support for the conclusions of \cite{haldane}, this underscores the necessity of economic firewalls for the prevention of systemic risk \cite{cl}.

Following the mean field scenario, where the wealthiest trader is the only survivor, it is natural to ask
if this would also hold when the range of interactions is limited. Somewhat surprisingly, this
turns out not always to be the case, with nontrivial and counter-intuitive survivor patterns being found often \cite{nirmal}. While
 one can certainly rule out the survival of a trader whose initial wealth is less than threshold
  ($X_{\star}$),  many-body interactions can then give rise to extremely
 complex dynamics in Stage II for  traders with  $X>X_{\star}$. This
points to the existence of ways of winning against the odds to (b)eat the competition, when interactions are limited in range; in the remainder of this paper, we use selective networking as a strategy to achieve this aim.

\section{Traders in complex networks\label{sec_compnet}}

In this section, we examine the mechanisms of selective networking: a particularly interesting example to consider is the class of small-world networks. These have the property that  long- and short-range interactions can coexist;  such networks also contain `hubs', where certain sites are preferentially endowed with many connections. Small-world networks can be constructed by starting with regular lattices, adding links randomly with  probability $p$ to their sites \cite{watts} and then freezing them, so that the average degree of the sites is increased for all $p>0$.

\begin{figure}[!t]
\begin{center}
\begin{tabular}{cc}
\hspace{-0.4cm}\resizebox{70mm}{60mm}{\includegraphics{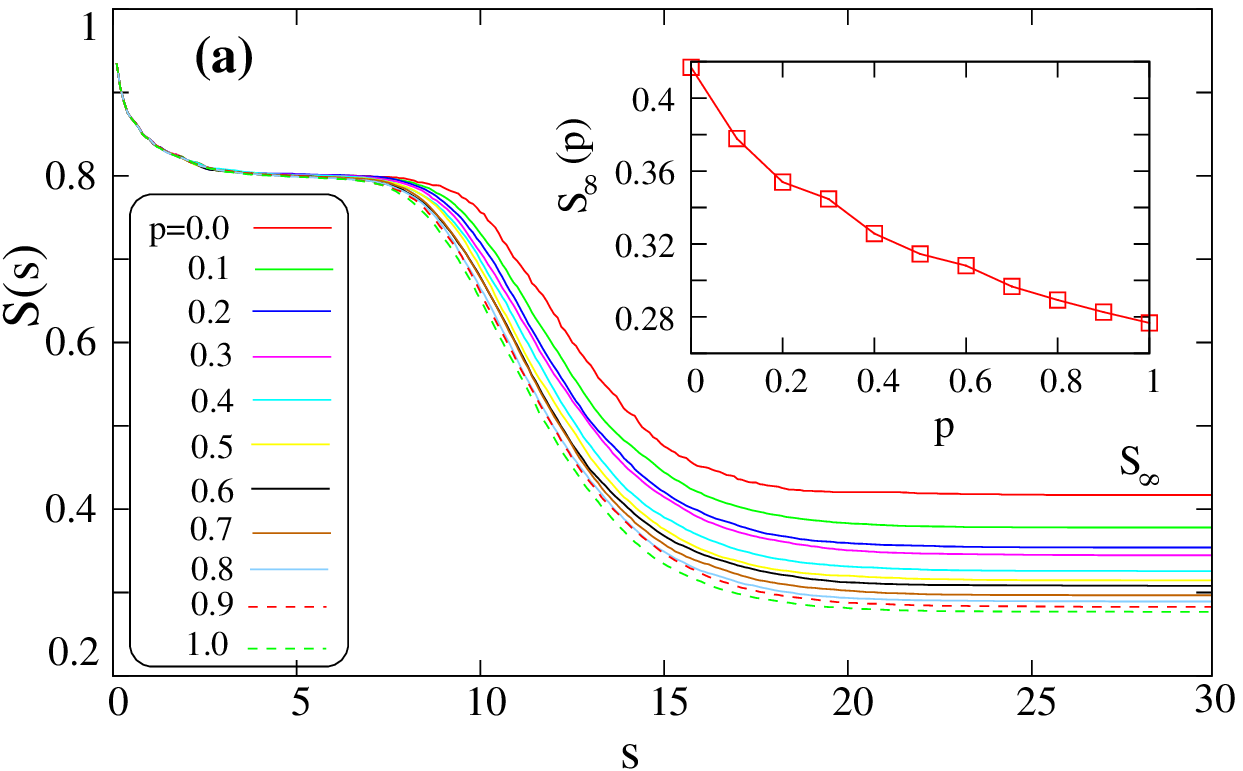}}&
\hspace{0cm}\resizebox{70mm}{60mm}{\includegraphics{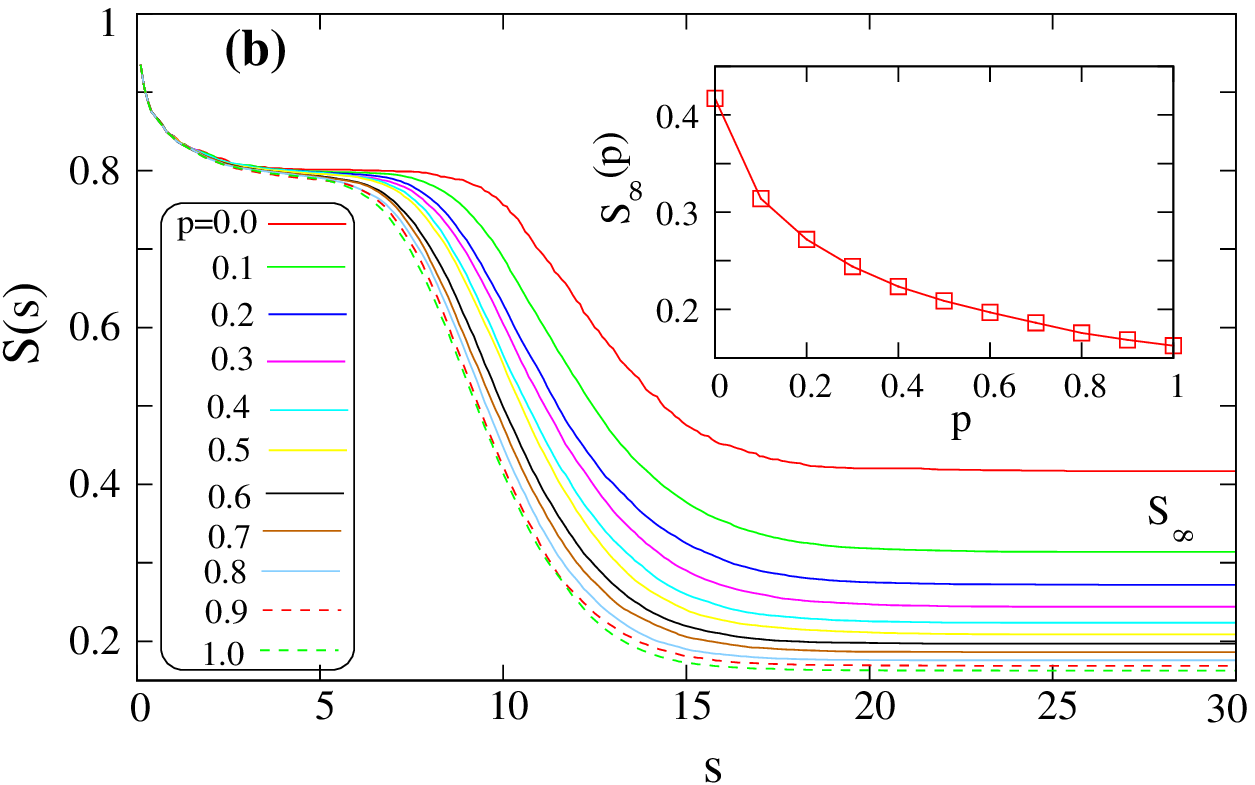}}\\
\end{tabular}

\begin{tabular}{cc}
\hspace{-0.4cm}\resizebox{70mm}{60mm}{\includegraphics{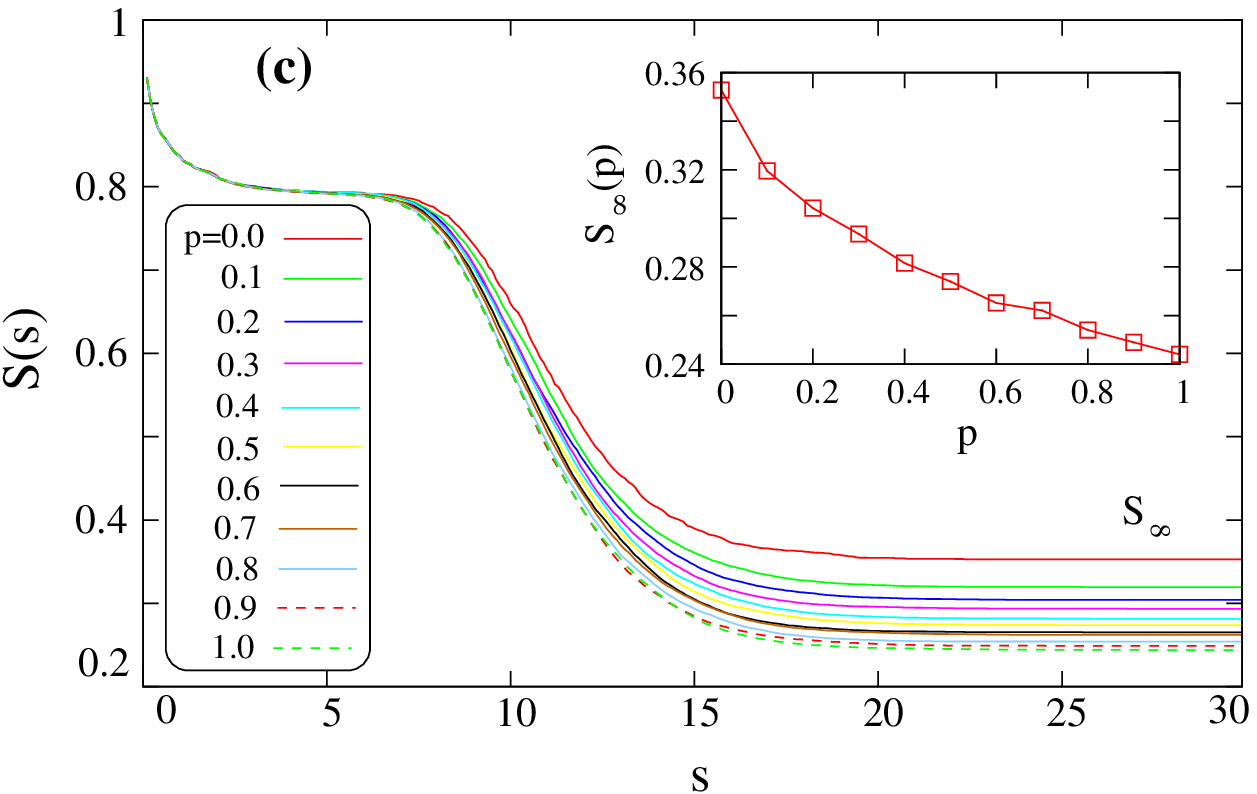}}&
\hspace{0cm}\resizebox{70mm}{60mm}{\includegraphics{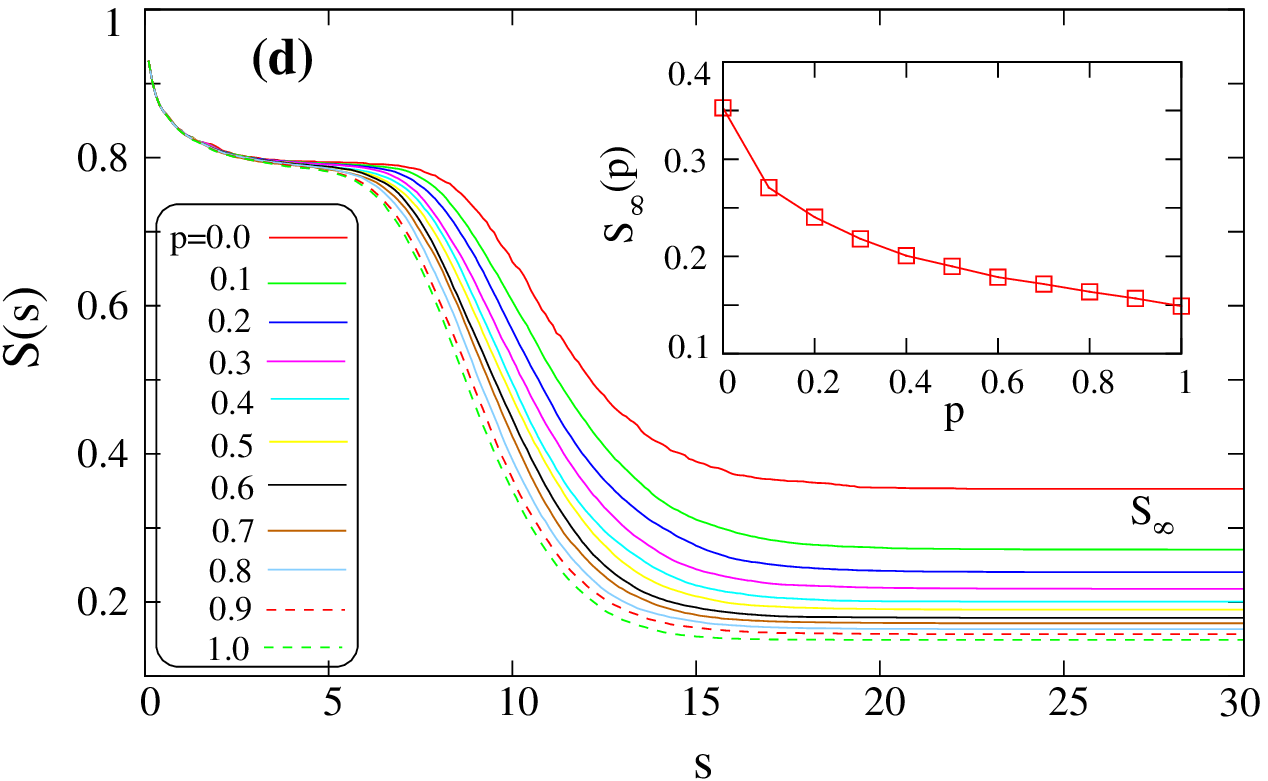}}\\
\end{tabular}
\end{center}
\caption{ (Color online) The plots of $S(s)$  as a function of reduced time $s$, for increasing values of wiring probabilities $p$. The  $1$-cycle scheme is shown in (a) for a $1$-dimensional ring and (c) for a $2$-dimensional square lattice. The $5$-cycle scheme is shown in (b) for a $1$-dimensional ring and (d)  for a $2$-dimensional square lattice.  The insets in all the figures show the asymptotic survival ratios $S_{\infty}$ as a function of the probability $p$. Here, the system size for the $1$-d ring is  $2000$ and for the $2$-d square lattice  it is $50 \times 50$ - all our data is averaged  over $10$ random network configurations.\label{one}}
\end{figure}

\begin{figure}[!t]
\begin{center}
\resizebox{90mm}{75mm}{\includegraphics{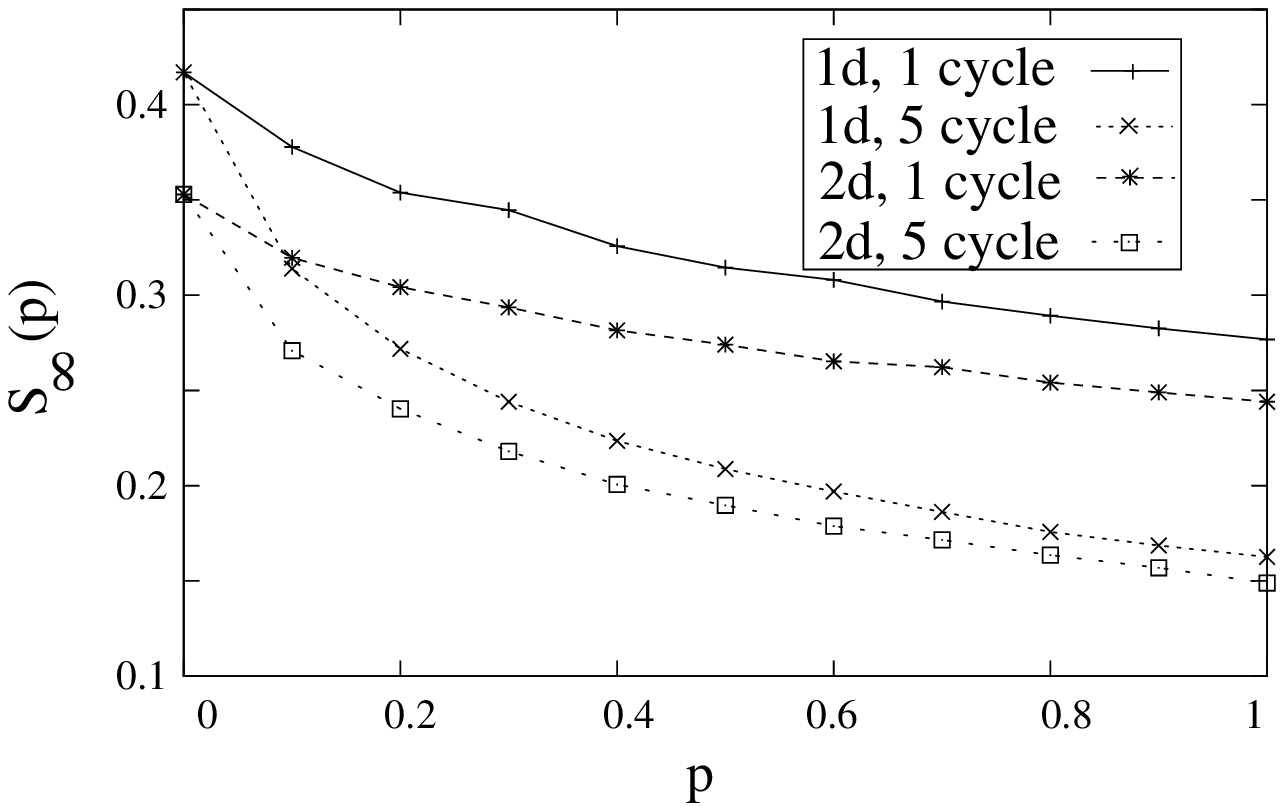}}\\
\end{center}
\caption{The plot of asymptotic survival ratios $S_{\infty} (s)$ as a function of the probabilities $p$ for $1$- and $5$-cycle schemes in $1$  and $2$ dimensions. The error bars for all the graphs do not exceed $0.003$ and are smaller than the plot symbols.\label{asymp_all}}
\end{figure}

\subsection{One-dimensional ring and two-dimensional square lattices}

Consider a regular one-dimensional ring lattice of size $N=2000$. To start with, the wealth of the traders located on the lattice sites evolve according to  Eqn. \ref{finite_model},  where the interactions are  with nearest neighbours only. Next, the lattice is modified by adding new links between sites chosen randomly with an associated probability $p$. For $p=0$, the network is  ordered, while for $p=1$, the network becomes completely random. 

In the  first scheme \cite{nirmal}, we add links probabilistically starting with site $i=1$ and end with $i=N$, only once: we call this the $1$-cycle scheme.  The survival ratios of traders as a function of reduced time $s$ for different values of wiring probability $p$ are presented in Fig. \ref{one}(a). Consider the $p=0$ case, which corresponds to a regular lattice; here the survival ratio $S(s)$ shows two stages,
Stage I  and Stage II, in its evolution. For all values of $0 < p \le 1$,  the existence of these well-separated Stages I and  II is also observed. There is a noticeable fall in the survivor ratio  as $p$ is increased, though; this is clearly visible in  the  asymptotic values $S_{\infty}(p)$ plotted with respect to $p$ in the inset of Fig. \ref{one}(a). As  $p$ increases, the number of links increases, leading to more interaction and competition between the traders, and hence  a  decrease in the number of survivors \cite{haldane}. As expected,  this behaviour interpolates between the two characteristic behaviours  relevant to the regular lattice and  mean field scenarios.

Next we implement the $5$-cycle scheme \cite{nirmal}, where the rewiring is done five times. Figure \ref{one}(b) shows the survivor ratio $S(s)$ as a function of $s$, where a clear decrease of $S(s)$ for increasing $p$. The asymptotic survival ratios $S_{\infty}$ for the $1$- and $5$-cycle schemes are shown in the insets of Fig. \ref{one}(a) and (b) respectively, with a clear decrease of $S_{\infty}$ as $p$ increases. In addition, the survivor ratio in the $5$-cycle scheme is  consistently smaller than in the $1$-cycle scheme, for all $p$.

The above procedures are repeated for  a two-dimensional square lattice of size $50\times50$ \cite{nirmal}, and  survival ratios obtained as a
 function of $s$ for  the $1$- and $5$-cycle schemes (see Figs. \ref{one}(c)-(d)). The asymptotic survivor ratios
for these two cases are shown in the insets of Figs. \ref{one}(c)-(d); they follow a decreasing trend with increasing $p$, similar to the $1d$ case. Finally, we plot the asymptotic survival ratios of the $1$- and $5$-cycle schemes in $1d$ and $2d$, in Fig. \ref{asymp_all}. 

All the above simulations reinforce one of the central themes of this paper, that economic firewalls are good ways to avoid systemic collapse in a predatory scenario, since the more globalised the interactions, the greater the systemic risk \cite{haldane}.

\section{Networking strategies: the Lazarus effect \label{networksmall}}

Since the central feature of this model is the survival of traders against the competition, it is of great interest to find a smart networking strategy which can change the fate of a trader, for example, by reviving a 'dying' trader  to life -- we call this the Lazarus \cite{luke} effect.

We systematically investigate  the effect of adding a finite number of non-local connections to a chosen central trader. In \cite{nirmal}, it was shown that  the growth or decay of the wealth of a trader is solely dictated by its relative rate of change versus the cumulative rate of change of its neighbours' wealth. The key to better survival should
therefore lie in choosing to network with traders whose wealth is decaying strongly. We accordingly divide all possible non-local connections into two classes:  class A comprises  eventual non-survivors ($X<X_{\star}$), while class B comprises would-be survivors ($X>X_{\star}$) ). In the next subsection, we look at the outcome of networking with members of class A. 

\subsection{Non-local connections with  eventual non-survivors ($X_i < X_{\star}$)}
Recall that non-survivors ($X<X_{\star}$) die very early during  Stage I. In connecting such
traders to a given trader with $X>X_{\star}$, we can be sure that they will never be able to
compete with him, much less run him out of business. 

\begin{figure}[!t]
\begin{center}
\begin{tabular}{cccc}
\hspace{-1cm}(a)&
\hspace{-0.4cm}\resizebox{50mm}{50mm}{\includegraphics{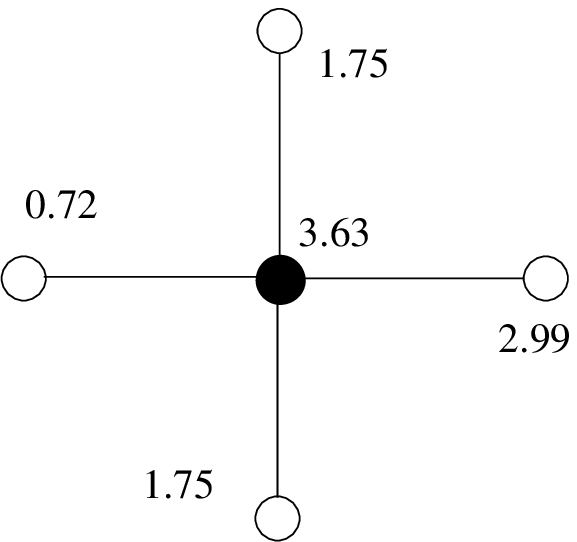}}&
\hspace{0cm}(b)&
\hspace{0.0cm}\resizebox{70mm}{60mm}{\includegraphics{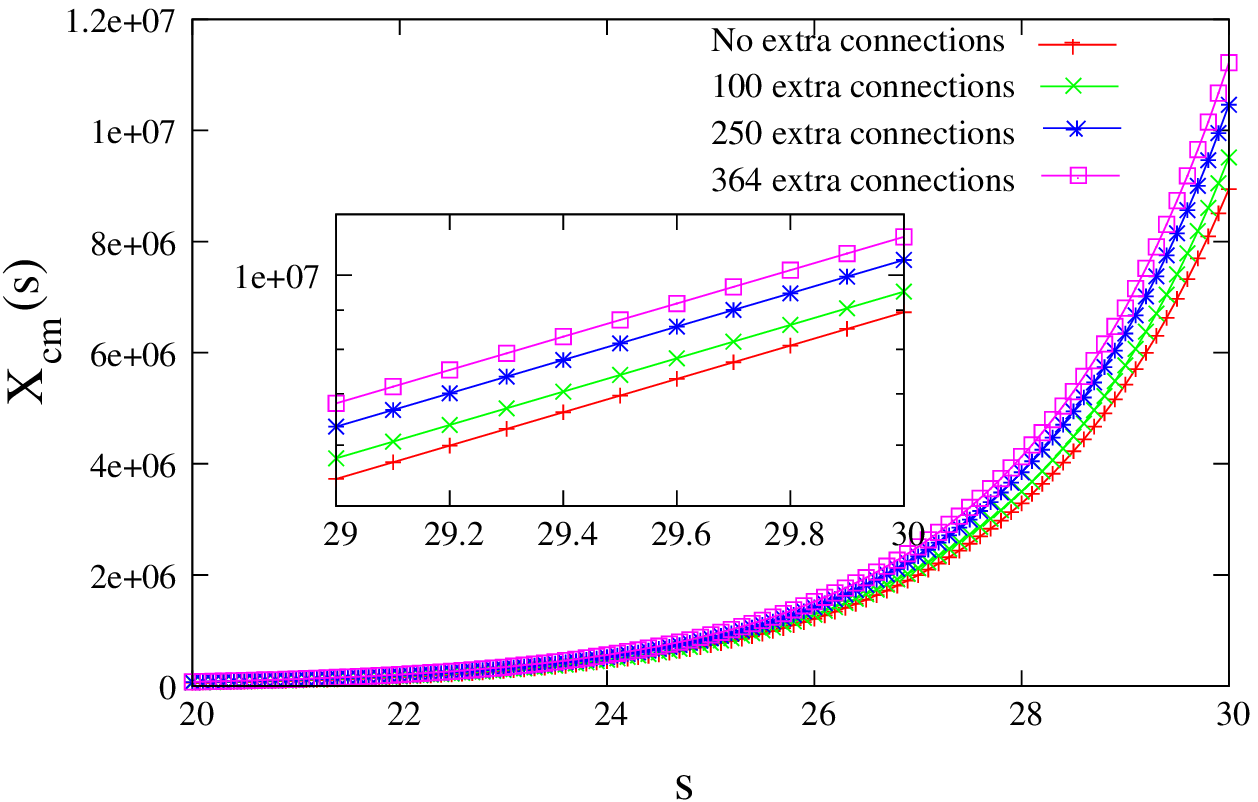}}\\
\end{tabular}

\end{center}
\caption{ (Color online) (a) The central trader here is a survivor in his original configuration. (b) The asymptotic wealth of the  central trader diverges with increasing connectivity to eventual non-survivors. \label{nonsurvive}}
\end{figure}

Let us consider a central trader who is an eventual survivor, as shown in Fig. \ref{nonsurvive}. We now let him network with eventual non-survivors from all over the lattice, and record the growth of his wealth as a function of the number of traders in his network; the results are shown in Fig. \ref{nonsurvive} (b). When all the neighbours go out of business, their contribution in Eqn. 3 is zero, leading to the exponential solution shown in Fig. \ref{nonsurvive} (b). The wealth of   the central trader increases markedly as more and more small traders are connected to him, making him an even wealthier survivor asymptotically.

\begin{figure}[!b]
\begin{center}
\begin{tabular}{cc}
\hspace{-1cm}(a)&
\hspace{-0.4cm}\resizebox{90mm}{70mm}{\includegraphics{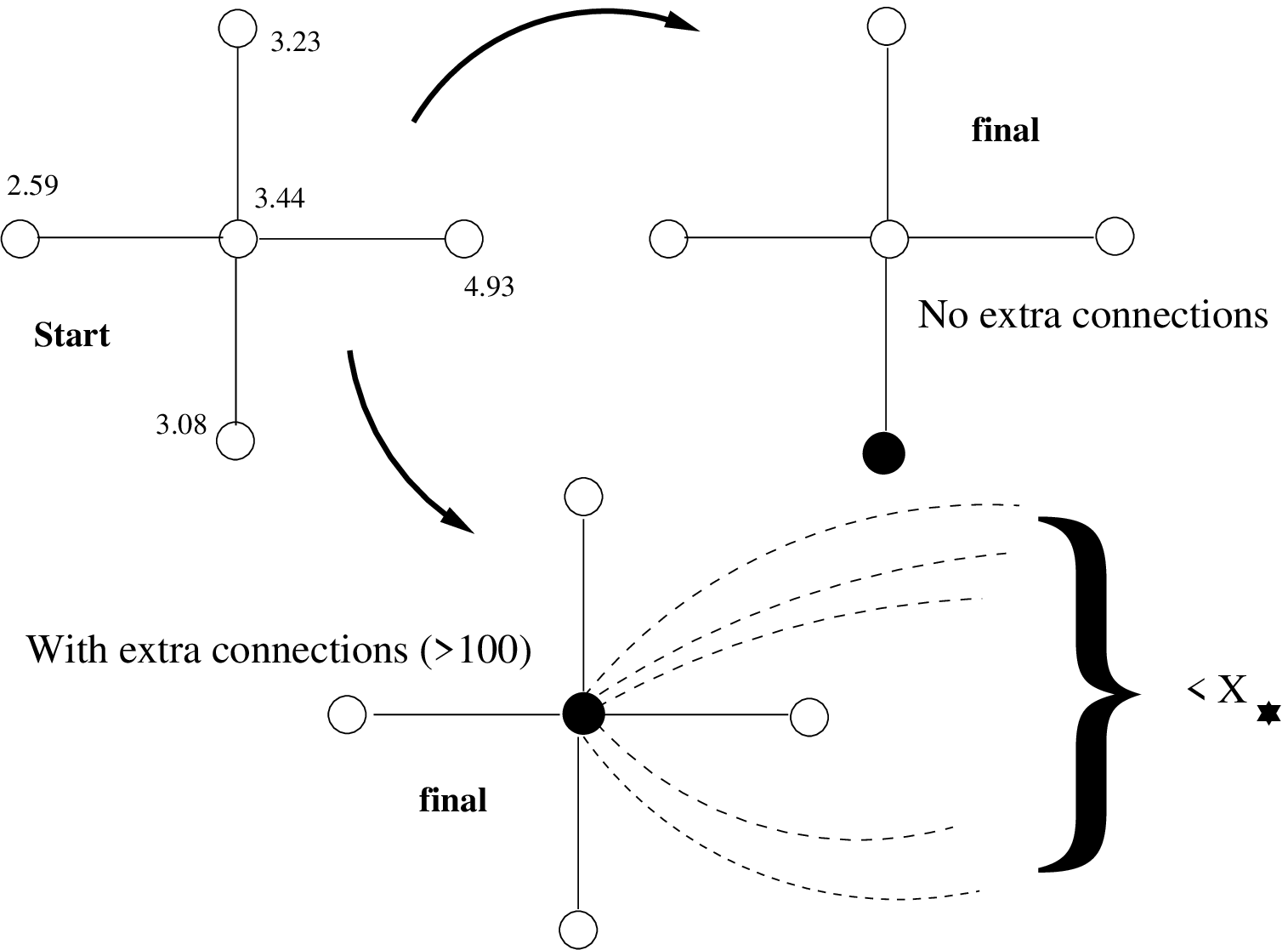}}\\
\hspace{-1cm}(b)&
\hspace{-0.4cm}\resizebox{90mm}{60mm}{\includegraphics{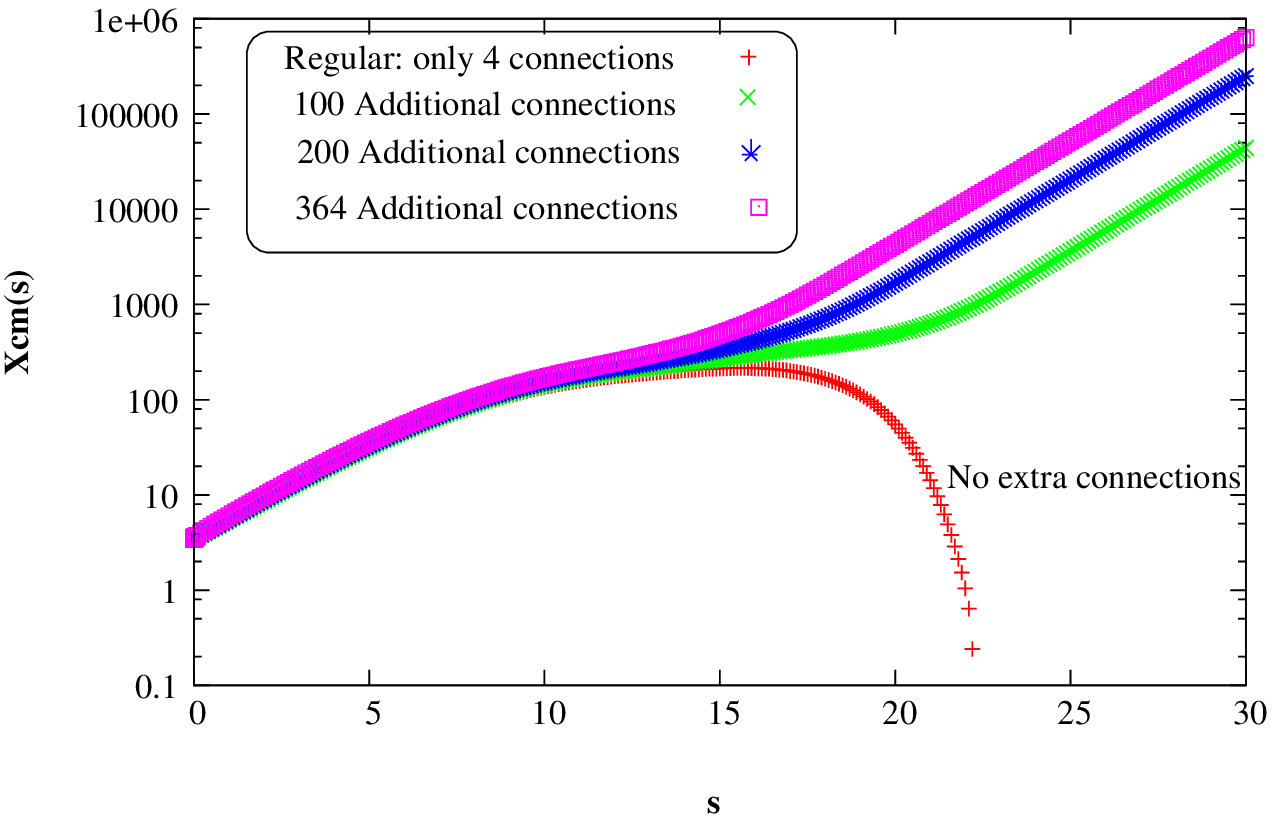}}\\
\end{tabular}

\end{center}
\caption{ (Color online) (a)  The central trader goes bankrupt in his original configuration without 
extra connections. He becomes a survivor after linking up with more and more non-survivors. (b) A crossover is seen here as the 
central trader is returned to (financial) 'life', his wealth $X_{cm}$ increasing with increasing  connections to non-surviving traders ($X < X_{\star}$).  \label{givelife}}
\end{figure}

We now use this observation to return a 'dying' trader to life. Figure \ref{givelife} shows our results:   the central trader would eventually have gone bankrupt in his original environment, but on  adding $100$ small traders  (whose wealth $X < X_{\star}$) to his network, he comes back to solvency: further additions , e.g.  $200$  or  $364$ such traders, evidently make him an even wealthier survivor (Fig. \ref{givelife} (b)). Thus, networking with eventual non-survivors is the surest way to invoke the Lazarus effect on a 'dying' trader.

\subsection{Networking with would-be survivors $X_i > X_{\star}$}

Choosing to network with traders whose intrinsic wealth is greater than $X_{\star}$ could turn out to be rather delicate. The financial 
lifespan of such traders will certainly exceed Stage I: and depending on their individual environments, they could either survive through
Stage II with a positive growth rate,  or die as a result of a negative growth rate. Networking with such traders is like playing Russian roulette.

\begin{figure}[!t]
\begin{center}
\resizebox{120mm}{100mm}{\includegraphics{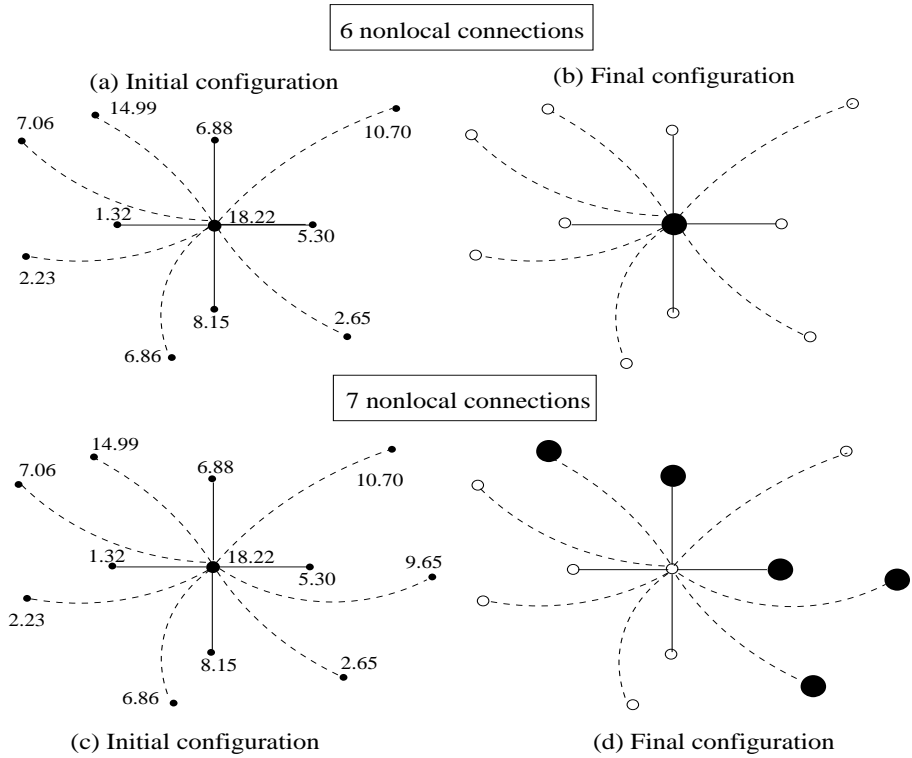}}\\
\end{center}
\caption{The central trader is networked with would-be survivors ($X_i>X_{\star}$) who are poorer than himself ($X_i<X_{cm}$).   (a) The initial configuration, where the central trader has $6$ non-local connections, (b) the asymptotic state with the only survivor being the central trader. (c) With the linkage of one more non-local trader to the existing configuration  (d) the central trader goes bankrupt. In (b) and (d), the open (dark) circles represent asymptotic non-survivors (survivors). \label{schem636}}
\end{figure}

Consider first  non-local connections with  would-be survivors ($X_i > X_{\star}$) who are
poorer than our chosen trader ($X_i < X_{cm}$). Such would-be survivors will live beyond Stage I, their wealth showing at least initially a positive growth
rate (Fig. \ref{freewealth}).  From a  mean field perspective,  we would therefore expect 
to see  a decrease in the chances of survival of the chosen trader  as it networks with more and more would-be survivors.

\begin{figure}[!b]
\begin{center}
\resizebox{120mm}{90mm}{\includegraphics{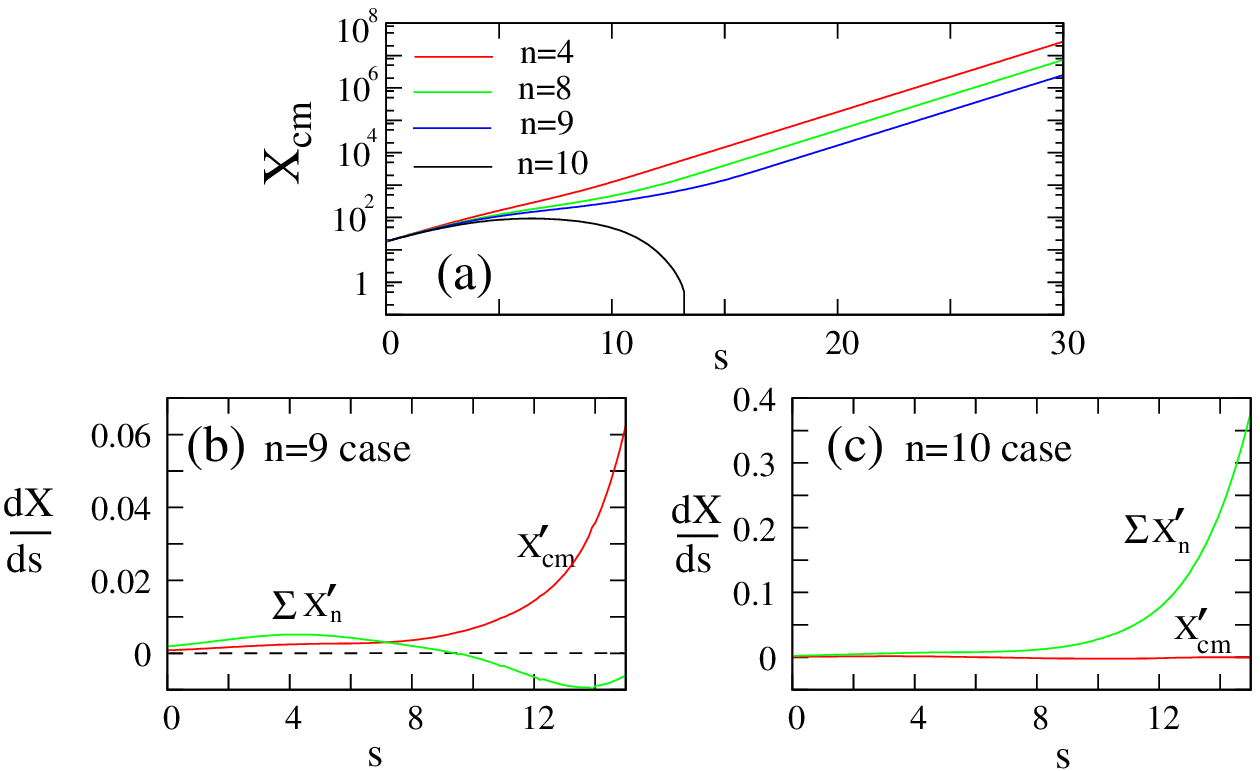}}\\
\end{center}
\caption{ (Color online)The central trader   is networked with traders whose $X_n > X_{\star}$. In (a) the growth of $X_{cm}$ with an increasing number of connections is plotted -  $n=4$ corresponds to a regular lattice. There is a crossover seen when the number of connections increases from  $n=10$ to $n=11$; for larger $n$ values, the central site goes bankrupt. This observation is supported by the  rates of growth of $X_{cm}$ and its neighbours $X_{n}$   (b) when $n=10$ and (c) when $n=11$. \label{636}}
\end{figure}

 Figure \ref{schem636} shows a sample scenario, where the central trader is connected non-locally with would-be survivors who are poorer than himself. In Fig. \ref{schem636} (a) the central trader networks with 6 would-be survivors and is able to survive asymptotically (Fig. \ref{schem636} (b) ). On the other hand, adding one more would-be survivor (Fig. \ref{schem636} (c)) to the existing network of the central trader, causes him to go out of business at long times (Fig. \ref{schem636} (d)). As the central trader  is made bankrupt by the arrival of the new connection, the fates of some of his other links are also changed (cf. Figs. \ref{schem636} (b) and (d)).

To understand the dynamics, we present the rates of growth  for another sample scenario in Fig. \ref{636}.
An  increase in the number of non-local connections  with would-be 
survivors, leads to a fall in the absolute value of $X_{cm}$ as well as its rate of growth $X'_{cm}$. Beyond
 a certain number of networked contacts, the wealth of the central trader begins to decay, and eventually vanishes. This crossover from life to death happens when the cumulative rate  of the wealth growth of the neighbours $\Sigma X'_{i,j}$  
 is larger than that of the central trader $X'_{cm}$. Unfortunately, however,  the intricate many-body nature of this problem precludes a prediction of when such crossovers might occur in general.

\begin{figure}[!t]
\begin{center}
\resizebox{120mm}{100mm}{\includegraphics{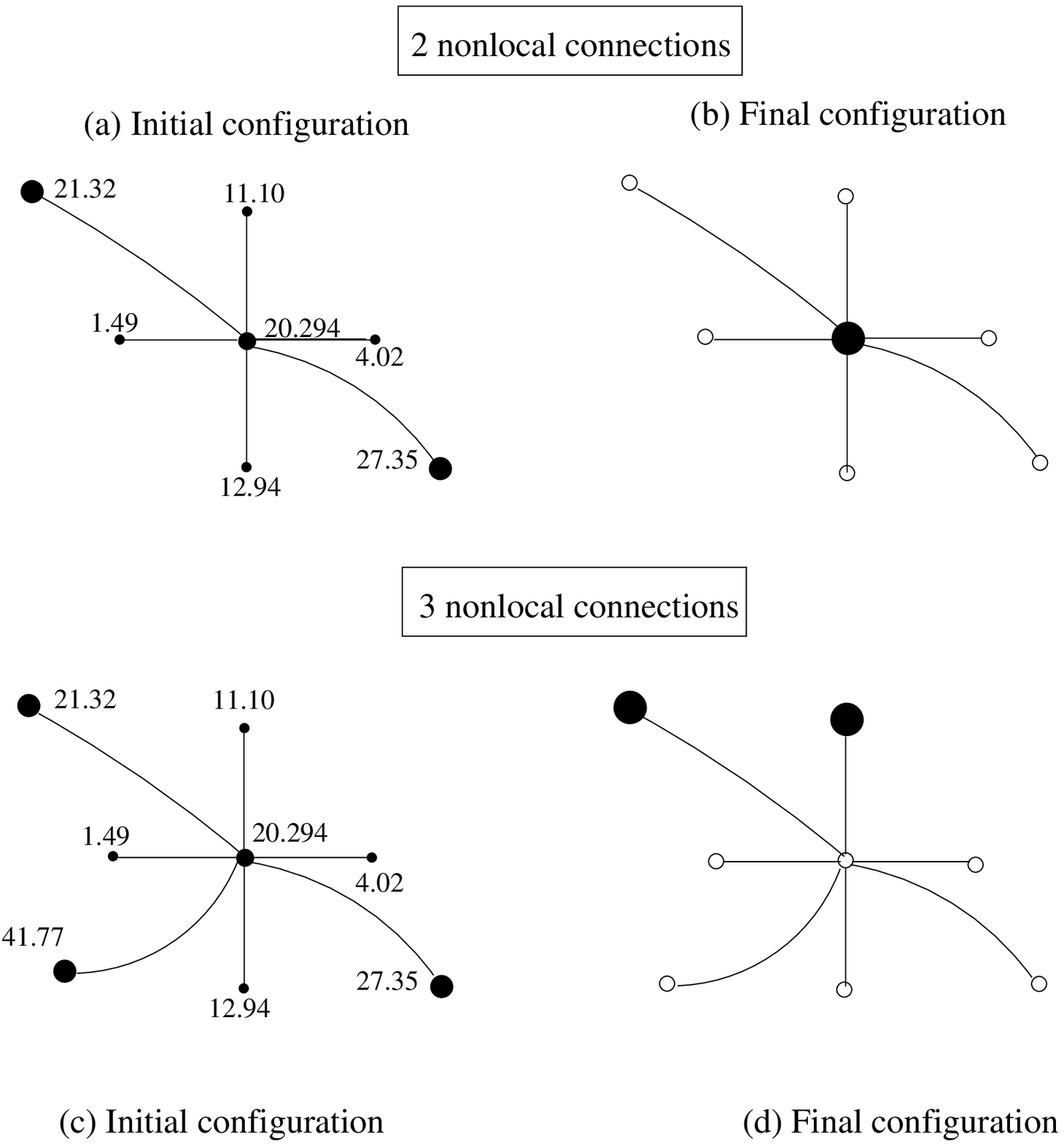}}\\
\end{center}
\caption{The central trader is networked with would-be survivors ($X_i>X_{\star}$) that are wealthier than himself($X_i > X_{cm}$).   (a) The initial configuration with $2$ non-local connections, (b) and its asymptotic state; the only survivor here is the central trader. (c) With the addition of one more non-local trader to the existing configuration (d) the central trader goes bankrupt.  In (b) and (d), the open (dark) circles represent asymptotic non-survivors (survivors).\label{schem1054}}
\end{figure}

\begin{figure}[!b]
\begin{center}
\resizebox{120mm}{90mm}{\includegraphics{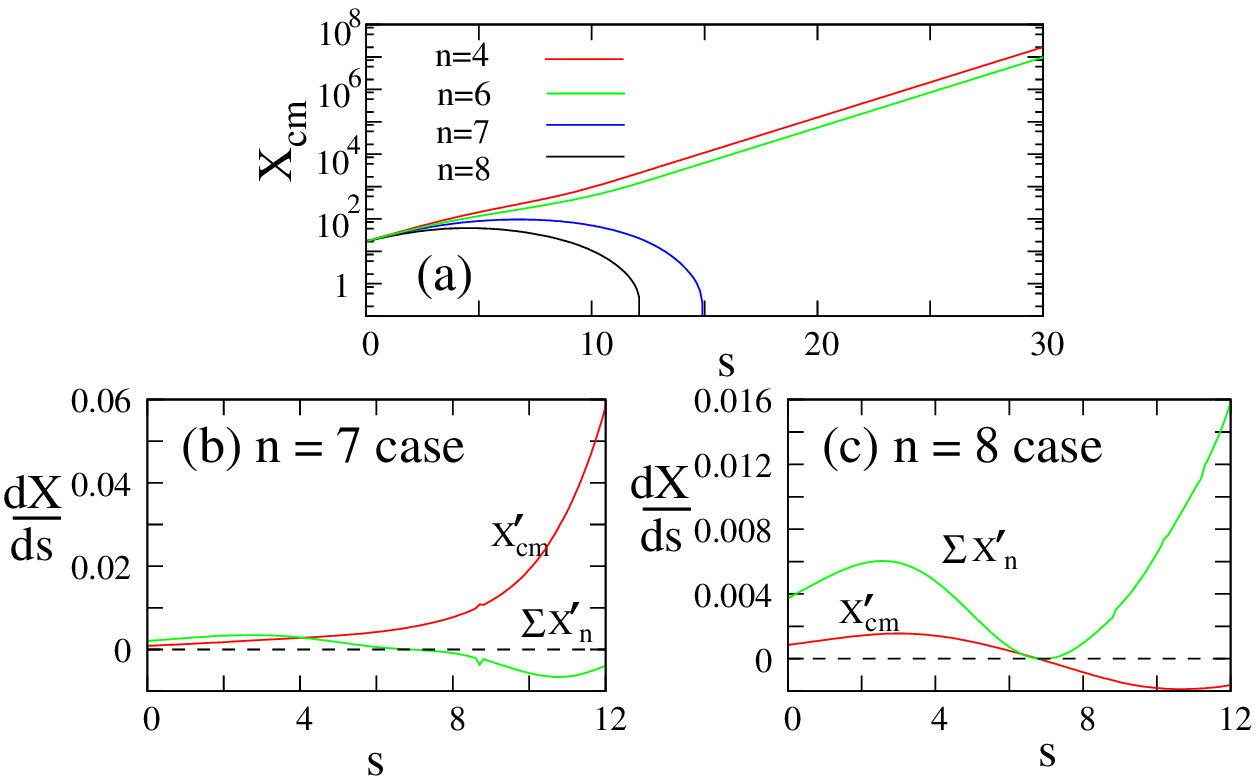}}\\
\end{center}
\caption{ (Color online) The central trader of a configuration given in Fig. \ref{schem1054}(b) is connected with non-local traders with $X_n > X_{cm}$. In (a) the growth of $X_{cm}$ with an increasing number of connections is plotted -  $n=4$ corresponds to a regular lattice. There is a crossover seen when the number of connections increases from  $n=6$ to $n=7$; for larger $n$ values, the central trader goes bankrupt. This observation is supported by the  rates of growth of the wealth of the central trader, $X_{cm}$, and that of its neighbours $X_{n}$,   (b) when $n=6$ and (c) when $n=7$. \label{1054}}
\end{figure}

 Finally, in the case where a given  trader networks with would-be survivors who are richer than himself ($X_i > X_{\star}$ and $X_i > X_{cm}$), one would expect a speedier 'death'. One such sample scenario is depicted in Fig. \ref{schem1054} and the corresponding rates of evolution of the traders' wealth in Fig. \ref{1054}.  We notice that in his original configuration with four neighbours ($n=4$)), the central trader is a survivor. As we increase the number of networked connections, the growth rate of his wealth gets stunted;  there is a 
substantial fall for 2 extra links ($n=6$ in Fig. \ref{1054}). Adding one more link ($n=7$) does the final damage; the central trader goes bankrupt.  The rates shown in Fig. \ref{1054} (b) and (c) for $n=6$  and $n=7$ connections
vividly capture the competition for survival, leading to solvency in one case and bankruptcy in the other.

As expected, we observe that  fewer connections (here, $n=7$) are needed, compared to the earlier case with smaller would-be survivors ($n=11$), to eliminate the chosen trader. In closing, we should of course emphasise that the $n$ values mentioned here are illustrative.

\section{Survivor distributions and rare events \label{univ}}

We have seen that the safest strategy for the Lazarus effect is to network with eventual non-survivors, i.e. those who will never get past Stage I. It is also relatively safe to network with would-be survivors, provided they are poorer than oneself. (This would explain why multinationals are not afraid to enter an arena where smaller-size retailers predominate, for example). However, in this section we consider rare events, where the wealthiest trader in a given neighbourhood dies marginally, and a poorer one survives against the odds.

 We look first at 
the four  immediate neighbours  of a given trader, and consider their pairwise interactions with him. Clearly, had such a pair been isolated, the larger trader would have won \cite{luck_am}. However, many-body interactions in the lattice mean that this is not always true. We therefore ask the question:   what is the proportion of  cases  where the poorer trader wins?  
\begin{figure}[!t]
\begin{center}
\resizebox{120mm}{90mm}{\includegraphics{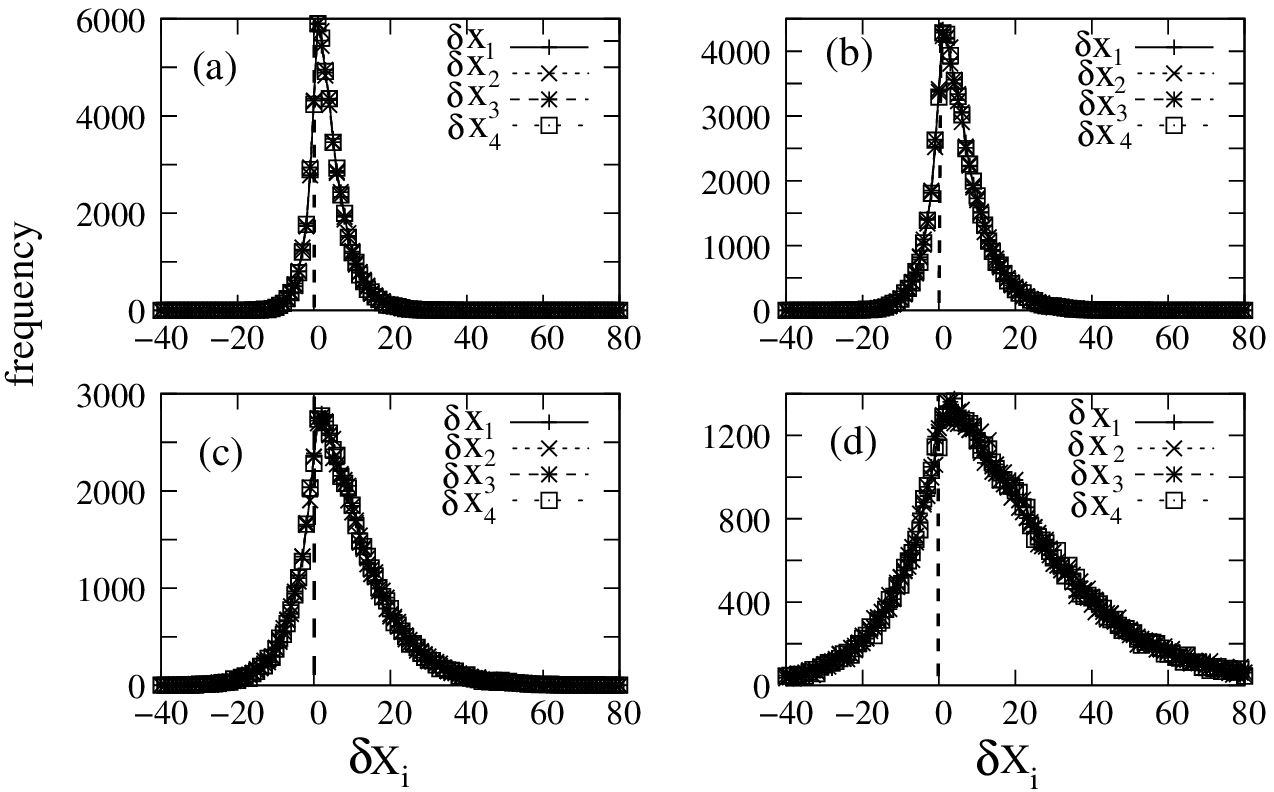}}\\
\end{center}
\caption{The plots show the distribution of pairwise wealth differences between  survivors and their four neighbours. They are obtained for  exponential distributions of initial wealth with  different mean values   $1/\mu$, where $\mu = -\log(S_1)/X_{\star}$.  The plots are for (a) $1/\mu = 3.92$ ($S_1 = 0.6$) (b) 5.607  ($S_1 = 0.7$) (c) 8.963  ($S_1 = 0.8$) and (d) 18.982  ($S_1 = 0.9$). The system size is $400 \times 400$. \label{univ_fig}}
\end{figure}

\begin{figure}[!t]
\begin{center}
\resizebox{80mm}{60mm}{\includegraphics{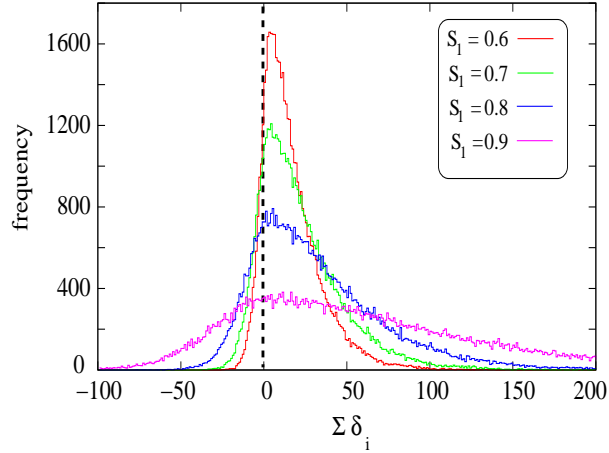}}\\
\end{center}
\caption{(Color online) The plots show the distribution of wealth differences between survivors and all of their four neighbours. They are obtained for  exponential distributions of initial wealth with different mean values  $1/\mu$, where $\mu = -\log(S_1)/X_{\star}$.  The plot in  `red' represents the case for  $1/\mu = 3.92$ ($S_1 = 0.6$), `green' represents $1/\mu = 5.607$  ($S_1 = 0.7$), `blue' represents $1/\mu = 8.963$  ($S_1 = 0.8$), and  `pink' represents $1/\mu = 18.982$  ($S_1 = 0.9$). The system size is $400 \times 400$. \label{univ_cuml}}
\end{figure}

\begin{figure}[!t]
\begin{center}
\resizebox{120mm}{90mm}{\includegraphics{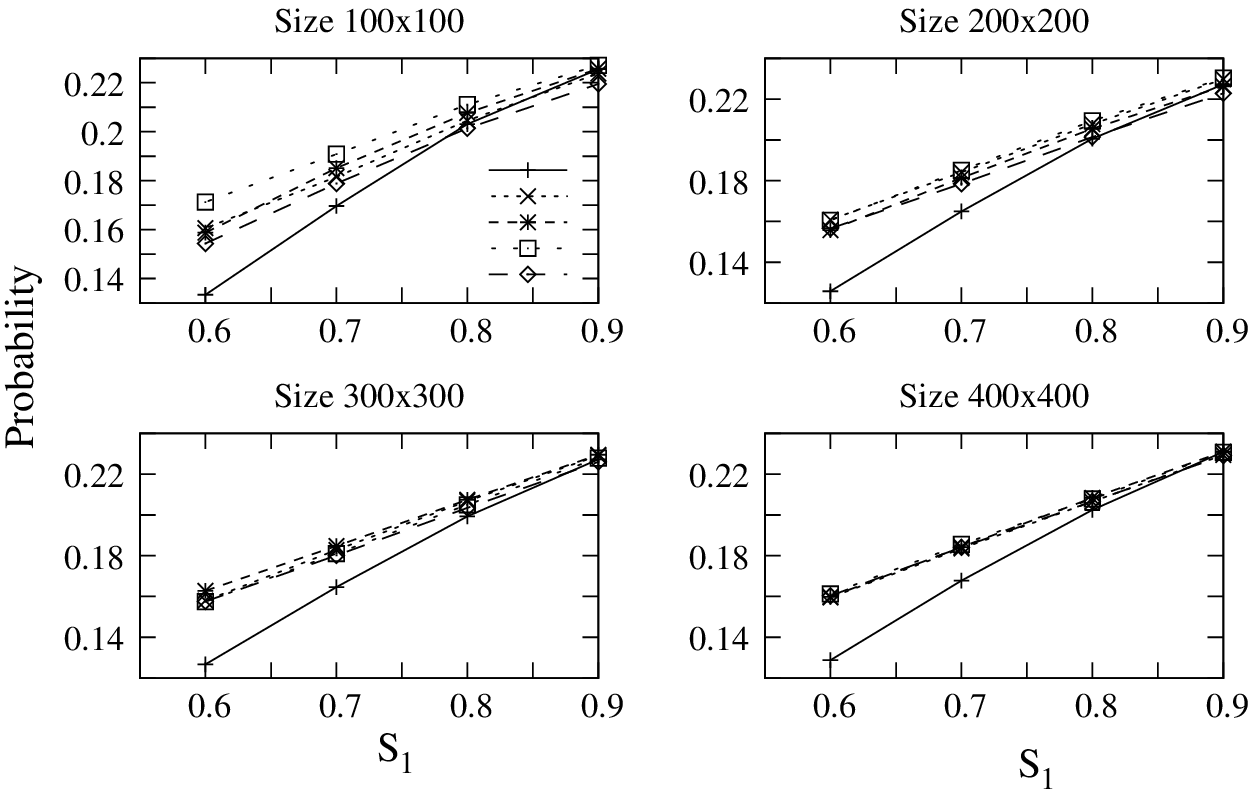}}\\
\end{center}
\caption{The plots show the  probability of finding a poorer survivor, who has won against the odds. The probability increases with increase in $S_1$ (refer to Figs. \ref{univ_fig} and \ref{univ_cuml}). While the solid line shows the combined cumulative probabilities (CCP) (from Fig. \ref{univ_cuml}), the rest of the lines represent the individual cumulative probabilities (from Fig. \ref{univ_fig}). \label{all_prob}}
\end{figure}

Each survivor has four neighbours; we first calculate the probability distribution of the initial wealth differences in a pairwise fashion between a survivor and each of his neighbours. The initial wealth differences are given by $\delta X_i = X_{cm} - X_i$ ($i=1,2,3,4$)  corresponding to the four neighbours - right, left, bottom and top - of a survivor. The distribution of $\delta X_i$ for all the survivors is shown in Fig. \ref{univ_fig}. Here, a negative $\delta X_i$ means that the survivor  is poorer than his neighbour, and conversely for positive $\delta X_i$. All four  distributions corresponding to  four neighbouring pairs  overlap due to isotropy; the resulting distributions are universal functions of {\it wealth differences}, depending  only on $\mu$ .

We also obtain the cumulative wealth difference between survivors and all of their four neighbours  viz. $4 X_{cm} - \sum_{i=1}^4 X_i = \sum_{i=1}^4 \delta X_i$ (see Fig. \ref{univ_cuml}). The distributions of $\sum_{i=1}^4 \delta_i$  are plotted in Fig. \ref{univ_cuml} for different values of $\mu$. For a positive cumulative wealth difference we know that the survivor is richer than his neighbours, matching our intuition based on the mean-field regime. The  negative side of the distribution is more interesting, {\it comprising traders who are poorer than their four neighbours combined, and who have won against the odds}.

 Notice that  both the survivor-neighbour pair distribution (Fig. \ref{univ_fig}), and the survivor- all neighbours cumulative distribution (Fig. \ref{univ_cuml}) get broader with increasing $\mu$.  This is because increasing $\mu = - \log(S_1)/X_{\star}$ \cite{luck_am} increases the number of potential survivors $S_1$ beyond Stage 1.   In each case,  the fraction of area under the negative side of the survivor pair-distribution
 gives an estimate of  survivors against the odds -- an example of  some of the rare events alluded to at the beginning of this section.
   
  Figure \ref{all_prob} shows this fraction, both in terms of  individual survivor-pair distributions and cumulative distributions, as a function of the 
$\mu$ of the initial wealth distribution, for different system sizes. 
There are more survivors, hence more survivors
\textit{against} the odds, leading to an increase in the fraction plotted on the y-axis of
Fig. \ref{all_prob} for both  distributions. For the largest
system size, there is full isotropy in the pairwise distributions; the probability of finding a survivor
against the odds
 is now  seen to be a {\it regular and
universal function} of  $\mu$ in both pairwise and cumulative cases, relying only on wealth differences
rather than on wealth. Finally, the cumulative distribution gives a more stringent survival criterion than the pairwise one, as is to be expected from the global nature of the dynamics.

A major conclusion to be drawn from Fig. \ref{all_prob} is the following: there are
traders who \textit{are eliminated} against the odds (traders who are wealthier than the eventual survivor). These should be easier to revive (as they have failed marginally) by selective networking than those who have failed because they are indeed worse off. This question is of real economic relevance, and its mathematical resolution seems to us to be an important open problem.

\section{Discussion \label{discuss}}

We have used the model of \cite{luck_am} to investigate two related issues in this paper on predatory trading \cite{am_wealth}: first, that of systemic risk in the presence of increasing interactions, and next, the use of selective networking to prevent financial collapse. As long-range connections are introduced with  probability $0<p<1$ to individual traders \cite{nirmal}, we find that the qualitative features of the networked system remain the same as that of the regular  case. The presence of two well-separated dynamical stages is retained, and the glassy dynamics and metastable states of \cite{luck_am} persist. However,  the number of survivors decreases as expected with increasing $p$, quantitatively validating the thesis if \cite{haldane}; and systemic risk is far greater as the complexity of interactions is increased. This view finds resonance with the present economic scenario, where it appears that some measure of insulation via economic firewalls, is needed to prevent individual, and hence eventually systemic, collapse.

Another central result of this paper is the use of smart networking strategies  to modify the fate of an arbitrary trader. We find that it is safest to network with eventual non-survivors; their decay and eventual death lead to the transformation of the destiny of a given site, from bankruptcy to solvency, or from solvency to greater solvency. Networking with peers, or with those who are born richer, in general leads to the weakening of one's own finances, and an almost inevitable bankruptcy, given a predatory scenario.

However, the above is not immutable:  the  probability distributions in the last section of the paper indicate an interesting universality of survival `against the odds'.  It would be interesting to find a predictive way of financial networking that would enable such a phenomenon to occur both at the individual, and at the societal, level.

\begin{acknowledgement}
AM gratefully acknowledges the input of J M Luck, A S Majumdar and N N Thyagu to this manuscript.
\end{acknowledgement}
%


\end{document}